\documentclass[aps,prl,twocolumn,showpacs,floatfix,superscriptaddress]{revtex4}
\usepackage{amsmath,amssymb,amsfonts,bbm,graphicx}
\usepackage[dvips,usenames]{color}

\def\Tr{\hbox{Tr}}

\def\Fth{F_{\rm e}}
\def\psith{\psi_{\rm e}}
\def\bmsigma{\boldsymbol{\sigma}}
\def\bmSigma{\boldsymbol{\Sigma}}
\def\bmX{\boldsymbol{X}}
\def\bmR{\boldsymbol{R}}
\def\bmA{\boldsymbol{\Sigma}}
\def\bmB{\boldsymbol{\Sigma}}
\def\bmC{\boldsymbol{\Sigma}}

\newtheorem{theorem}{Theorem}
\newtheorem{corollary}{Corollary}

\newcommand{\qed}{\hfill $\Box$\medskip}

\begin{document}
\title{Fidelity matters: the birth of entanglement in the 
mixing of Gaussian states}
\author{Stefano Olivares}\email{stefano.olivares@ts.infn.it}
\affiliation{Dipartimento di Fisica, Universit\`a degli Studi di
Trieste, I-34151 Trieste, Italy}
\affiliation{CNISM UdR Milano Statale, I-20133 Milano, Italy}
\author{Matteo G.~A.~Paris}\email{matteo.paris@fisica.unimi.it}
\affiliation{Dipartimento di Fisica, Universit\`a degli Studi di 
Milano, I-20133 Milano, Italy}
\affiliation{CNISM UdR Milano Statale, I-20133 Milano, Italy}
\date{\today}
\begin{abstract}
  We address the interaction of two Gaussian states through bilinear
  exchange Hamiltonians and analyze the correlations exhibited by the
  resulting bipartite systems. We demonstrate that entanglement arises
  if and only if the fidelity between the two input Gaussian states
  falls under a threshold value depending only on their purities,
  first moments and on the strength of the coupling.  Our result
  clarifies the role of quantum fluctuations (squeezing) as a
  prerequisite for entanglement generation and provides a tool to
  optimize the generation of entanglement in linear systems of
  interest for quantum technology.
\end{abstract}
\pacs{03.67.Mn}
\maketitle
Gaussian states (GS), that is quantum states with Gaussian Wigner functions,
play a leading role in continuous variable quantum technology
\cite{revs} for their extremal properties \cite{wlf06} and because they
may be generated with current technology, in particular in the quantum
optics context \cite{gra03,dau:09,dae:10}.  As a consequence, much
attention have been dedicated to the characterization of Gaussian
entanglement
\cite{sim:00,dua:00,gio03,pvl03,ser04,ade:04,hyl06,sch06,mar:08}. Among
the possible mechanisms to generate Gaussian entanglement, the one
consisting in mixing squeezed states
\cite{par97,wan02,kim02,zhu04,ade:06,nha09,spe:11} is of special
interest in view of its feasibility, which indeed had been crucial to
achieve continuous variable teleportation \cite{fur:98}. The entangling
power of bilinear interactions has been widely analyzed, either to
optimize the generation of entanglement \cite{par:99,wol:03} or to find
relations between their entanglement and purities \cite{ade:04b} or
teleportation fidelity \cite{pir:03,ade:05}.
\par
In this Letter we address bilinear, energy conserving, i.e., exchange,
interactions described by Hamiltonians of the form $H_I= g (a^\dag b + a
b^\dag)$, where $a$ and $b$ are bosonic annihilation operators,
$[a,a^\dag]=1$ and $[b,b^\dag]=1$, and $g$ the coupling constant.
Hamiltonians of this kind are suitable to describe very different kinds
of quantum systems, such as, e.g., two light-mode in a beam splitter or
a frequency converter, collective modes in a gases of cold atoms
\cite{meystre}, atom-light nondemolition measurements \cite{Nat11}, 
optomechanical oscillators \cite{pir:03,xia:10},
nanomechanical oscillators \cite{cav08},
and
superconducting resonators \cite{chi:10}, all of which are of 
interest for the quantum technology. Our analysis can be applied to all
these systems and lead to very general results about the resources
needed for Gaussian entanglement generation.  
\par  
The bilinear Hamiltonians $H_I$ generally describe the action of simple
passive interactions and, in view of this simplicity, their fundamental
quantum properties are often overlooked. Actually, the exchange
amplitudes for the quanta of one of the systems strongly depend on the
statistics of the quanta of the other one and on the particle
indistinguishability. This mechanism gives rise to interference and,
thus, to the birth of correlations in the output bipartite system. A
question arises about the nature of these correlations, depending on the
parameters of the input signals and coupling constant. In this Letter,
motivated by recent results on the dynamics of bipartite GS
through bilinear interactions \cite{kim:09,oli:09} and by their
experimental demonstration \cite{blo:10}, we investigate the relation
between the properties of two input GS and the correlations
exhibited by the output state. Our main result is that entanglement
arises if and only if the fidelity between the two input states falls
under a threshold value depending only on their purities, first-moment
values and on the strength of the coupling. Our analysis provides a
direct link between the mismatch in the quantum properties of the input
signals and the creation of entanglement, thus providing a better
understanding of the process leading to the generation of nonclassical
correlations. In fact, if, from the one hand, it is well known that
squeezing is a necessary resource to create entanglement
\cite{wan02,kim02,wol:03}, from the other hand in this Letter we show
what is the actual role played by the squeezing, that is making the two
input GS different enough to entangle the output system.
\par  
The most general single-mode Gaussian state can be written as $\varrho =
\varrho(\alpha,\xi,N) = D(\alpha)S(\xi) \nu_{\rm
th}(N)S^{\dag}(\xi)D^{\dag}(\alpha)$, where $S(r) = \exp[\frac12
(\xi{a^{\dag}}^2-\xi^*a^2)]$ and $D(\alpha) = \exp[ \alpha a^{\dag} -
\alpha^*a)]$ are the squeezing operator and the displacement operator,
respectively, and $\nu_{\rm th}(N) = (N)^{a^\dag a}/(1+N)^{a^\dag a +
1}$ is a thermal equilibrium state with $N$ average number of quanta,
$a$ being the annihilation operator. Up to introducing the vector of
operator $\bmR^T=(R_1,R_2)\equiv(q,p)$, where $q=(a+a^\dag)/{\sqrt{2}}$
and $p=(a-a^\dag)/(i\sqrt{2})$ are the so-called quadrature operators,
we can fully characterize $\varrho$ by means of the first moment vector
$\overline{\bmX}^T = \langle\bmR^T \rangle = \sqrt{2}(\Re{\rm
e}[\alpha],\Im{\rm m}[\alpha])$, with $\langle A \rangle = \Tr[A\,
\varrho]$, and of the $2\times 2$ covariance matrix (CM) $\bmsigma$, with
$[\bmsigma]_{hk}=\frac12\langle R_hR_k + R_kR_h \rangle - \langle
R_h\rangle\langle R_k \rangle$, $k=1,2$, which explicitly reads:
$[\bmsigma]_{kk} = (2\mu)^{-1}\left[ \cosh(2r)
-(-1)^k\cos(\psi)\sinh(2r) \right]$ for $k=1,2$ and $ [\bmsigma]_{12} =
[\bmsigma]_{21} = -(2\mu)^{-1} \sin(\psi)\sinh(2r)$, where we put $\xi =
r e^{i\psi}$, $r,\psi \in {\mathbbm R}$ and introduced the purity of the
state $\mu=\Tr[\varrho^2]=(1+2N)^{-1}$. Since we are interested in the
dynamics of the correlations, which are not affected by the first
moment, we start addressing GS with zero first moments
($\alpha = 0$). The general case will be considered later on in this
Letter.
\par
When two uncorrelated, single-mode GS $\varrho_k$ with CMs $\bmsigma_k$,
$k=1,2$, interacts through the bilinear Hamiltonian $H_I$, the $4\times
4$ CM $\bmSigma$ of the evolved bipartite state $\varrho_{12} = U_g(t)\,
\varrho_1\otimes \varrho_{2}\,U^{\dag}_g(t)$, $U_g(t) = \exp\{- i H_I
t\}$ being the evolution operator, can be written in the following
block-matrix form \cite{revs}: 
\begin{equation}
\bmSigma = \left(\begin{array}{cc}
\bmA_1 & \bmC_{12}\\ [1ex]
\bmC_{12} & \bmB_2
\end{array}\right),\quad
\begin{array}{l}
\bmA_1 = \tau\bmsigma_1 + (1-\tau)\bmsigma_2, \\
\bmB_2 = \tau\bmsigma_2 + (1-\tau)\bmsigma_1 ,\\
\bmC_{12} = \tau(1-\tau)(\bmsigma_2-\bmsigma_1),
\end{array}\label{evCM}
\end{equation}
$\tau = \cos^2(gt)$ being an effective coupling parameter, and where
the presence of a nonzero covariance term $\bmC_{12}$ suggests the
emergence of quantum or classical correlations between the two
systems.  Since $\bmC_{12}$ depends on the difference between the
input states CMs, a question naturally arises about the relation
between the ``similarity'' of the inputs and the birth of (nonlocal)
correlations. In this Letter we answer this question and demonstrate
that entanglement arises if and only if the fidelity between the two
input GS falls under a threshold value, which depends only on their
purities, the value of the first moments, and the coupling $\tau$.
\par
Let us now consider the pair of uncorrelated, single-mode GS
$\varrho_k=\varrho(\xi_k,N_k)$, $k=1,2$ and assume, without loss of
generality, $\xi_1 = r_1$ and $\xi_2 = r_2 e^{i\psi}$, with $r_k,\psi
\in {\mathbbm R}$. After the interaction, we found that the presence
of entanglement at the output is governed by the sole fidelity
$F(\varrho_1,\varrho_2) =
[\Tr(\sqrt{\sqrt{\varrho_1}\,\varrho_2\sqrt{\varrho_1}})]^2$ between
the inputs. Our results may be summarized by the following:
\begin{theorem}\label{theo:fid}
  The state $\varrho_{12} = U_g(t)\, \varrho_1\otimes
  \varrho_{2}\,U^{\dag}_g(t)$, resulting from the mixing of two
  GS  with zero first moments,   $\varrho_1(r_1,N_1)$ and
  $\varrho_2(r_2 e^{i\psi},N_2)$, is   entangled if and only if the
  fidelity $F(\varrho_1,\varrho_2)$   between the inputs falls below a
  threshold value   $\Fth(\mu_1,\mu_2;\tau)$, which depends only on
  their purities   $\mu_k = \Tr[\varrho_k^2] = (1+2N_k)^{-1}$, $k=1,2$,
  and on the effective coupling parameter $\tau =   \cos^2(gt)$.
\end{theorem}
\begin{proof}
In order to prove the theorem, we recall that a bipartite Gaussian
state
$\varrho_{12}$ is entangled if and only if the minimum symplectic
eigenvalue $\tilde\lambda$ of CM associated with the partially
transposed state is $\tilde\lambda < 1/2$ \cite{sim:00}. Moreover,
without loss of generality, we can address the scenario in which $r_k$
and $N_k$, $k=1,2$, are fixed and we let $\psi$ vary in the interval
$[0,2\pi]$.  
\par
\begin{figure}[h!]
\includegraphics[width=0.77\columnwidth]{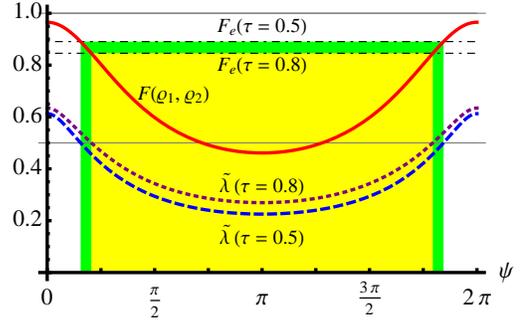}
\vspace{-0.3cm}
\caption{\label{f:FlvPsi} (Color online) Plot of the fidelity
$F(\varrho_1,\varrho_2)$ (red, solid line) between the two input
states and of the minimum symplectic eigenvalue $\tilde\lambda$ as a
function of $\psi$ for $\tau=0.5$ (blue, dashed
line) and $\tau=0.8$ (purple, dotted line). The other involved
parameters are $\xi_1=0.5$, $N_1=0.2$, $\xi_2 = 0.7 e^{i\psi}$ and
$N_2=0.3$.  The colored regions denote the ranges of $\psi$ leading to an 
entangled state for the given $\tau$, while the horizontal dot-dashed 
lines refer to the corresponding thresholds $\Fth$.}
\end{figure}
First of all, we prove that $\tilde\lambda < 1/2 \Rightarrow
F(\varrho_1,\varrho_2) < \Fth(\mu_1,\mu_2;\tau)$ (necessary condition).
As we will see, this will allow us to find the analytic expression of
the threshold $\Fth(\mu_1,\mu_2;\tau)$, which will be used to prove the
sufficient condition, i.e., $F(\varrho_1,\varrho_2) <
\Fth(\mu_1,\mu_2;\tau) \Rightarrow \tilde\lambda < 1/2$.
Fig.~\ref{f:FlvPsi} shows the typical behavior of $\tilde\lambda$ and of
the fidelity $F$ as a function of the squeezing phase $\psi$ for fixed
$r_k$ and $N_k$, $k=1,2$ (here we do not report their analytic
expressions since they are quite cumbersome).  As one can see, both
$\tilde\lambda$ and $F$ are monotone, decreasing (increasing) functions
of $\psi$ in the interval $[0,\pi)$ ($[\pi,2\pi]$, respectively) and
have a minimum at $\psi=\pi$, whose actual value depends on both $r_k$
and $N_k$ but not on $\tau$. In our case, one finds that, for fixed
$r_k$ and $N_k$, $k=1,2$, if $\min_{\psi}\tilde\lambda < 1/2$, then
there exists a threshold value $\psith \equiv \psith
(r_1,\mu_1,r_2,\mu_2,\tau)$:
\begin{align*}
\psith = \arccos\left\{
\frac{\cosh(2r_1)\cosh(2r_2)-f(\mu_1,\mu_2,\tau)}{\sinh(2r_1)\sinh(2r_2)}
\right\},
\end{align*}
where we introduced:
\begin{align*}
f(\mu_1,\mu_2,\tau) =
\frac{1+\mu_1^2\mu_2^2-(\mu_1^2+\mu_2^2)(1-2\tau)^2}
{8\,\mu_1\mu_2\tau(1-\tau)},
\end{align*}
and $\mu_k = \Tr[\varrho_k^2] = (1+2N_k)^{-1}$, $k=1,2$, are the
purities of the inputs, such that if $\psi\in(\psith,2\pi -\psith)$ then
$\tilde\lambda < 1/2$, i.e., $\varrho_{12}$ is entangled. 
Since the fidelity between the two GS $\varrho_k$,
characterized by the CMs $\bmsigma_k$, $k=1,2$ (and zero first
moments), is given by \cite{scu:98} $ F(\varrho_1,\varrho_2) = \left(
  \sqrt{\Delta +\delta} - \sqrt{\delta}\right)^{-1}$, where $\Delta =
\det[\bmsigma_1 + \bmsigma_2]$ and $\delta = 4\prod_{k=1}^{2}
(\det[\bmsigma_k]-\frac14)$, the threshold value $\Fth \equiv
\Fth(\mu_1,\mu_2; \tau)$ of the fidelity is thus obtained by setting
$\psi = \psith$ and explicitly reads:
\begin{equation}
\label{F:th}
\Fth = \frac{4\mu_1\mu_2\sqrt{\tau(1-\tau)}}
{\sqrt{g_{-}+4 \tau(1-\tau) g_{+}}-\sqrt{4 \tau(1-\tau) g_{-}}},
\end{equation}
where $g_{\pm}\equiv g_{\pm}(\mu_1,\mu_2)=\prod_{k=1,2}(1\pm\mu_k^2)$.
The threshold depends only on $\tau$ and on the purities $\mu_k$ of
the input GS and is independent of the squeezing parameters $r_k$,
despite the fact $\psith$ does. Finally, if $\tilde\lambda < 1/2$,
i.e., $\varrho_{12}$ is entangled, then $F(\varrho_1,\varrho_2) <
\Fth(\mu_1,\mu_2;\tau)$. This concludes the first part of the proof.
\par
Now we focus on the sufficient condition, i.e.,
$F(\varrho_1,\varrho_2) < \Fth(\mu_1,\mu_2;\tau) \Rightarrow
\tilde\lambda < 1/2$. Thanks to the first part of the theorem and
since both $F$ and $\tilde\lambda$ are continuous functions of $\psi$,
for fixed $r_k$ and $N_k$, $k=1,2$, which have a minimum in
$\psi=\pi$, it is enough to show that $F_{\rm min}\equiv \min_{\psi}
F(\varrho_1,\varrho_2) < \Fth(\mu_1,\mu_2;\tau) \Rightarrow
\lambda_{\rm min} \equiv \min_{\psi} \tilde\lambda < 1/2$.  We have:
\begin{align*}
F_{\min} =
\frac{2 \mu_1\mu_2}
{\sqrt{1+\mu_1^2 \mu_2^2 + 2 \mu_1 \mu_2 \cosh[2(r_1+r_2)]} 
-\sqrt{g_{-}}},
\end{align*}
where $g_{-}$ is the same as in Eq.~(\ref{F:th}), and:
\begin{align*}
\tilde\lambda_{\rm min} = \frac12
\frac{\left[\gamma - \sqrt{\gamma^2- (2
      \mu_1\mu_2)^2}\right]^{\frac12}}{\sqrt{2}\mu_1\mu_2},
\end{align*}
with $\gamma = (\mu_1^2+\mu_2^2)(1-2\tau)^2 + 8\mu_1\mu_2 \tau (1-\tau)
\cosh[2(r_1+r_2)]$, respectively.  The inequality $F_{\rm   min} <
\Fth(\mu_1,\mu_2;\tau)$, where $\Fth(\mu_1,\mu_2;\tau)$ is given in
Eq.~(\ref{F:th}), is satisfied if $\gamma > 1 + \mu_1^2\mu_2^2$, which
leads to $\tilde\lambda_{\rm min} < 1/2$, as one may verify after a
straightforward calculation. Now, since $\tilde\lambda$ is a continuous
function of $\psi$, there exists a range of values centered at
$\psi=\pi$, where the minimum occurs, in which $\tilde\lambda < 1/2$
and, thus, $F(\varrho_1,\varrho_2) < \Fth(\mu_1,\mu_2;\tau)$, because of
the first part of the theorem (necessary condition). This concludes the
proof of the Theorem. 
\qed \end{proof}
\par
As a matter of fact, the presence of nonzero first moments does not
affect the nonclassical correlations exhibited by a bipartite Gaussian
state, which only depend on the CM \cite{revs}.  Thus, we can state the
following straightforward:
\begin{corollary}\label{cor:fid}
  If $\overline{\bmX}_k^{T} = \Tr[(q_k,p_k)\,\varrho_k] \ne 0$, where
  $q_k=(a_k+a_k^\dag)/{\sqrt{2}}$ and $p=(a_k^\dag-a_k)/(i\sqrt{2})$
  are the quadrature operators of the system $k=1,2$, then the state
  $\varrho_{12} = U_g(t)\, \varrho_1\otimes \varrho_{2}\,
  U^{\dag}_g(t)$ is entangled if and only if:
\begin{equation}\label{th2}
F(\varrho_1,\varrho_2) < \Gamma(\overline{\bmX}_1,
\overline{\bmX}_2)\,\Fth(\mu_1,\mu_2;\tau),
\end{equation}
where $\Fth(\mu_1,\mu_2;\tau)$ is still given in Eq.~(\ref{F:th}) and:
\begin{equation}
\Gamma(\overline{\bmX}_1,\overline{\bmX}_2) = \exp\left[ -
\mbox{$\frac12$}\,\overline{\bmX}_{12}^{T} (\bmsigma_1+\bmsigma_2)^{-1}
\overline{\bmX}_{12} \right],
\end{equation}
where $\overline{\bmX}_{12} = (\overline{\bmX}_1-\overline{\bmX}_2)$.
\end{corollary}
\begin{proof}
The proof follows from Theorem~\ref{theo:fid} by noting
that the presence of nonzero first moments does not modify the
evolution of the CM, whereas the expression of the fidelity 
becomes \cite{scu:98} $F(\varrho_1,\varrho_2) =
  \Gamma(\overline{\bmX}_1,\overline{\bmX}_2) 
  \left(\sqrt{\Delta +\delta} - \sqrt{\delta}\right)^{-1}\!\!$, 
  where $\Delta$ and $\delta$ have been defined above.
\qed
\end{proof}
\par
Theorem ~\ref{theo:fid} states that if the two Gaussian inputs are
``too similar'' the correlations induced by the interaction are local,
i.e., may be mimicked by local operations performed on each of the
systems.  The extreme case corresponds to mix a pair of identical GS:
in this case the interaction produces no effect, since the output
state is identical to the input one \cite{kim:09,oli:09}, i.e., a
factorized state made of two copies of the same input states, and we
have no correlations at all at the output. Notice that for pure (zero
mean) states the threshold on fidelity reduces to $F_{\rm e}
(1,1;\tau)=1$ $\forall \tau$, namely, any pair of not identical (zero
mean) pure GS gives raise to entanglement at the output. On the
contrary, two thermal states $\nu_k \equiv \nu_{\rm th}(N_k)$,
$k=1,2$, as inputs, i.e., the most classical GS, lead to
$F(\nu_{1},\nu_{2})>F_{\rm e}(\mu_{1},\mu_{2};\tau)$: this fact,
thanks to the Theorem~\ref{theo:fid}, shows that we need to squeeze
one or both of the classical inputs in order to make the states
different enough to give rise to entanglement. Notice, finally, that
the thresholds in Eqs.~(\ref{F:th}) and (\ref{th2}) involve strict
inequalities and when the fidelity between the inputs is exactly equal
to the threshold the output state is separable.
\par
The threshold $\Fth(\mu_1,\mu_2; \tau)$ is symmetric under the
exchange $\mu_1 \leftrightarrow \mu_2$ and if one of the two state is
pure, i.e., if $\mu_k=1$ then $F_{\rm e}(\tau) = \sqrt{2}\, \mu_h
/\sqrt{1+\mu_h^2}$, with $h\ne k$, i.e., the threshold no longer
depends on $\tau$.
\par 
For what concerns Gaussian entanglement, i.e., the resource
characterized by the violation of Simon's condition on CM
\cite{sim:00}, our results also apply to the case of non-Gaussian
input signals, upon evaluating the fidelity between the GS with the
same CMs of the non-Gaussian ones.  In fact, violation of the Simon's
condition is governed only by the behavior of the CM independently on
the Gaussian character of the inputs states. On the other hand,
identical non-Gaussian states may give raise to entangled output, the
mixing of two single-photon states in quantum optical systems being
the paradigmatic example \cite{HOM:87}. In other words, the
entanglement raising from the mixing of two identical non-Gaussian
states cannot be detected by Simon's condition on CM.
\par
\begin{figure}[th]
\vspace{-0.11cm}
\includegraphics[width=0.77\columnwidth]{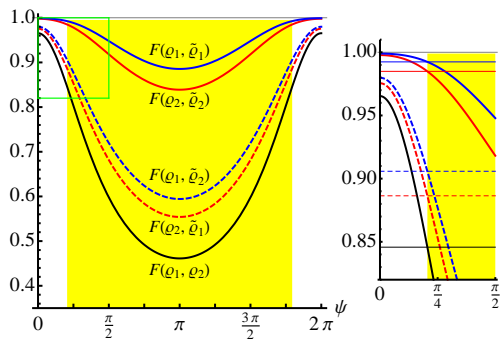}
\vspace{-0.11cm}
\caption{\label{f:InOut} (Color online) Plot of the fidelities
  $F(\varrho_h,\tilde\varrho_k)$ for $\tau=0.8$ and the same choice of
  the other involved parameters as in Fig.~\ref{f:FlvPsi}. The yellow
  region shows the interval of values of $\psi$ leading to an
  entangled state. The right panel is a magnification of the green,
  boxed region of the left panel: the horizontal lines refer to the
  corresponding thresholds $F_{\rm e}(\varrho_h,\tilde\varrho_k)$. See
  the text for details.}
\end{figure}

Up to now we have considered the correlation properties of the output
states with respect to the fidelity between the input ones. However,
similar relations may be found for the fidelities
$F(\varrho_h,\tilde\varrho_k)$, $k,h=1,2$, between the input and output
states, respectively, where $\tilde\varrho_h = \Tr_{k}[\varrho_{12}]$,
with $h\ne k$, are the reduced density matrices of the output states
taken separately. In this case we found that the output is entangled if
and only if $F(\varrho_h,\tilde\varrho_k) < F_{\rm
e}(\varrho_h,\tilde\varrho_k)$ where all the thresholds $F_{\rm
e}(\varrho_h,\tilde\varrho_k)$ still depends only on $\mu_1$, $\mu_2$
and $\tau$ (here we put as arguments the density matrices in order to
avoid confusion with the previous thresholds).  The analytic expressions
of $F_{\rm e}(\varrho_h,\tilde\varrho_k)$ are cumbersome and are not
reported explicitly, but we plot in Fig.~\ref{f:InOut} the input-output
fidelities and the corresponding thresholds for a particular choice of
the involved parameters. If we look at the interaction between the two
systems as a quantum noisy channel for one of the two, namely,
$\varrho_k \to {\cal E}(\varrho_k) \equiv \Tr_{h}[\varrho_{12}]$, $h\ne
k$, than the birth of the correlations between the outgoing systems
corresponds to a reduction of the input-output fidelity: the
correlations arise at the expense of the information contained in the
input signals. In turn, this result may be exploited for decoherence
control and preservation of entanglement using bath engineering \cite{blo:10}.  
\par
In conclusion, we have analyzed the correlations exhibited by two
initially uncorrelated GS which interact through a bilinear exchange
Hamiltonian. We found that entanglement arises if and only if the
fidelity between the two inputs falls under a threshold value
depending only on their purities, the first moments, and on the
coupling constant. Similar relations have been obtained for the
input-output fidelities. Our theorems clarify the role of squeezing as
a prerequisite to obtain entanglement out of bilinear interactions,
and provide a tool to optimize the generation of entanglement by
passive (energy conserving) devices.  Our results represent a progress
for the fundamental understanding of nonclassical correlations in
continuous variable systems and may found practical applications in
quantum technology.  Due to the recent advancement in the generation
and manipulation of GS, we foresee experimental implementations in
optomechanical and quantum optical systems.
\par
SO acknowledges support from the University of Trieste through the
``FRA 2009''.

\end{document}